\documentclass[11pt,showpacs,preprintnumbers,superscriptaddress,amsmath,amssymb,nofootinbib]{revtex4}
\usepackage{graphicx}
\usepackage{dcolumn}
\usepackage{bm}
\usepackage{amssymb}
\usepackage{amsmath}
\usepackage{epsfig}    
\usepackage{color}
\usepackage{slashed}
\usepackage{hhline}

\def\be{\begin{equation}}
\def\ee{\end{equation}}
\newcommand{\bea}{\begin{eqnarray}}
\newcommand{\eea}{\end{eqnarray}}
\newcommand{\nn}{\nonumber}

\numberwithin{equation}{section}

\begin{document}

\title{Inert Extension of the Zee-Babu Model}
\preprint{DCPT-14-146}
\preprint{IPPP-14-73}
\preprint{KIAS-P14049}
\author{Hiroshi Okada}
\email{hokada@kias.re.kr}
\affiliation{School of Physics, KIAS, Seoul 130-722, Korea}

\author{Takashi Toma}
\email{takashi.toma@durham.ac.uk}
\affiliation{Institute for Particle Physics Phenomenology\\
 University of
Durham, Durham DH1 3LE, United Kingdom}

\author{Kei Yagyu}
\email{keiyagyu@ncu.edu.tw}
\affiliation{Department of Physics and Center for Mathematics and Theoretical Physics,
National Central University, Chungli, Taiwan 32001, ROC}

\begin{abstract}

We propose a two-loop induced Zee-Babu type neutrino mass model at the TeV scale. 
Although there is no dark matter candidate in the original Zee-Babu model, 
that is contained in our model by introducing an unbroken discrete $Z_2$ symmetry. 
The discrepancy between the experimental value of the muon anomalous magnetic moment (muon $g-2$) and its prediction in the standard model 
can be explained by contributions from additional vector-like charged-leptons which are necessary to give non-zero neutrino masses. 
The mass of vector-like leptons to be slightly above 300 GeV is favored and allowed from the muon $g-2$ and the current LHC  data. 
We find that from the structure of neutrino mass matrix,  
doubly-charged scalar bosons in our model can mainly decay into the same-sign and same-flavour dilepton plus missing transverse momentum. 
By measuring an excess of these events at the LHC, our model can be distinguished from
the other models including doubly-charged scalar bosons. 

\end{abstract}
\maketitle
\newpage

\section{Introduction}

From the discovery of the Higgs boson at the CERN Large Hadron Collider (LHC)~\cite{Higgs_ATLAS,Higgs_CMS}, 
our standard picture for the spontaneous breakdown of electroweak symmetry has been confirmed. 
This fact tells us that the standard model (SM) well describes phenomena at collider experiments even including the Higgs sector.  

In spite of the great success of the SM, 
we need to consider new physics models beyond the SM, because 
there are phenomena which cannot be explained in the SM such as neutrino oscillations and the existence of dark matter (DM).
Therefore, further new particles are expected to be found at the 13 and 14 TeV runs of the LHC, which can be a direct evidence of new physics models. 

Radiative neutrino mass models give an attractive new physics scenario to explain tiny neutrino masses which 
are generated at loop levels.  
Because of loop suppression, masses of new particles 
can be taken at the TeV scale, so that radiative neutino mass models may be able to directly tested at collider experiments. 
The Zee model~\cite{Zee} and the Zee-Babu model~\cite{Zee-Babu} have been proposed in 1980s, in which 
Majorana type masses are generated at the one loop and two loop levels, respectively. 
After these models appeared, the three-loop neutrino mass model has been constructed by Krauss, Nasri and Trodden~\cite{Krauss:2002px} in the early 2000s, 
in which a right-handed neutrino running in the loop can be a DM candidate. 
In addition, the model by Ma~\cite{Ma:2006km} can explain neutrino masses at the one loop level with a DM candidate. 
The above two models provide the interesting connection between physics of neutrino and DM. 
Many other types of radiative neutrino mass models with DM have been considered in Refs.~\cite{
Aoki:2013gzs, Dasgupta:2013cwa, Aoki:2008av,
Schmidt:2012yg, Bouchand:2012dx, Aoki:2011he, Farzan:2012sa, 
Bonnet:2012kz, Kumericki:2012bf, Kumericki:2012bh, Ma:2012if, Gil:2012ya,
Okada:2012np, Hehn:2012kz, Dev:2012sg, Kajiyama:2012xg, Okada:2012sp,
Aoki:2010ib, Kanemura:2011vm, Lindner:2011it, Kanemura:2011mw,
Kanemura:2012rj, Gu:2007ug, Gu:2008zf, Gustafsson, Kajiyama:2013zla, Kajiyama:2013rla,
Hernandez:2013dta, Hernandez:2013hea, McDonald:2013hsa, 
Baek:2013fsa, Ma:2014cfa,  Ahriche:2014xra,
Kanemura:2011jj, Kanemura:2013qva, Kanemura:2014rpa,
Chen:2014ska, Ahriche:2014oda, Okada:2014vla, Ahriche:2014cda,
Aoki:2014cja, Lindner:2014oea},
and those with a non-Abelian discrete symmetry have been proposed in Refs.~\cite{Ahn:2012cg, Ma:2012ez,
Kajiyama:2013lja, Kajiyama:2013sza, Ma:2013mga, Ma:2014eka}. 
Recently, models which generate both masses of charged-leptons and neutrinos at the loop level have also been constructed in Refs.~\cite{radlepton1,radlepton2,radlepton3}. 
 
Among the various neutrino mass models, 
the Zee-Babu model mentioned in the above 
is constructed in a quite simple way, where isospin singlet singly- and doubly-charged scalar fields are added to the SM. 
However, this model does not have a DM candidate, because all the new particles are electromagnetically charged. 
In addition, it has been found that the lower bound on the masses of new charged scalar bosons are given between 1 and 2 TeV 
from the constraints of the most recent experimental data such as neutrino mixing with non-zero $\theta_{13}$\footnote{Analysis with 
the non-zero $\theta_{13}$ in the Zee-Babu model has also been performed in Refs.~\cite{Long1,Long2}.} 
and $\mu\to e\gamma$ with a perturbativity requirement, 
which makes difficult to directly discover these new particles at the LHC even with the 14 TeV collision energy~\cite{Schmidt:2014zoa}. 
Similar mass bounds have also been shown in Ref.~\cite{Nebot}, in which authors have also taken into account the recent Higgs boson search data at the LHC.

In this paper, we extend the Zee-Babu model so as to include a DM candidate by introducing a discrete $Z_2$ symmetry to the model\footnote{
The supersymmetric extension of the Zee-Babu model has been built in Ref.~\cite{Aoki:2010ib}, where a DM candidate is obtained as the lightest neutral R-parity odd particle.  }. 
In order to enclose the two-loop diagram, we need to add $Z_2$ odd fields; i.e., 
vector-like singly-charged leptons and an isospin doublet scalar field. 
A DM candidate is then obtained as the lightest neutral component of the doublet scalar field. 
In this model, the vector-like leptons play a crucial role not only to explain neutrino masses but also
the discrepancy between the experimental value of the muon anomalous magnetic moment (muon $g-2$) 
and its prediction in the SM~\cite{discrepancy1,discrepancy2}.   
We find that the mass of vector-like leptons to be slightly above 300 GeV is favored to explain the muon $g-2$, which is not excluded by 
the current experimental data. 

We then discuss the collider phenomenology, especially focusing on the production and decay of doubly-charged scalar bosons at the LHC.  
The doubly-charged scalar bosons can mainly decay into the same-sign dilepton with the same-flavour and missing transverse momentum due to the structure of neutrino mass matrix. 
This signal process is completely different with that from doubly-charged scalar bosons in the original Zee-Babu model and 
the Higgs triplet model (HTM)~\cite{Cheng:1980qt, Schechter:1980gr,
Lazarides:1980nt, Mohapatra:1980yp, Magg:1980ut} which contains isospin triplet scalar field, and one of its components corresponds to the doubly-charged state.

This paper is organized as follows.
In Sec.~II, we show our model building including the particle contents,
discussion of DM and the constraints from the oblique parameters. In Sec.~III, we calculate neutrino mass matrix and the muon $g-2$. 
Numerical evaluation of these observables is also given by using the current data for the neutrino masses and mixing. 
In Sec.~IV, the collider phenomenology of the new particles is
discussed. Summary and conclusion are given in Sec.~V.

\section{The Model}

\begin{center}
\begin{table}[t]
\begin{tabular}{c||c|c|c|c||c|c|c|c}\hline\hline  
&\multicolumn{4}{c||}{Lepton Fields} & \multicolumn{4}{c}{Scalar Fields} \\\hline
& ~$L_L^i$~ & ~$e_R^i$~ & ~$E_L^\alpha$~ & ~$E_R^\alpha$~ & ~$\Phi$~ &
 ~$\eta$~ & ~$\chi^+$~  & ~$\kappa^{++}$~ \\\hline 
$SU(2)_L$ & $\bm{2}$ & $\bm{1}$& $\bm{1}$ & $\bm{1}$ &
 $\bm{2}$&$\bm{2}$&$\bm{1}$&$\bm{1}$\\\hline 
$U(1)_Y$ & $-1/2$ & $-1$ & $-1$ & $-1$ & $+1/2$& $+1/2$& $+1$ & $+2$  \\\hline
$Z_2$ & $+$ & $+$ & $-$ & $-$ & $+$& $-$& $-$& $+$  \\\hline\hline
\end{tabular}
\caption{Contents of lepton and scalar fields
and their charge assignment under the $SU(2)_L\times U(1)_Y\times Z_2$ symmetry.
The indices $i$ (=1-3) and $\alpha$ (=1-3) denote the flavour for the SM leptons $L_L$ and $e_R$ and the new vector-like leptons $E_L$ and $E_R$, respectively. 
}
\label{tab:1}
\end{table}
\end{center}

\subsection{Lagrangian}

We extend the two-loop induced radiative neutrino mass model proposed by
Zee and Babu in Ref.~\cite{Zee-Babu} so as to 
contain a DM candidate by introducing an unbroken discrete $Z_2$ symmetry. 
The particle contents and the charge assignment are shown in Table~\ref{tab:1}. 
We add vector-like charged leptons $E_R^\alpha$ and $E_L^\alpha$ with three flavours; i.e., $\alpha=$1-3. 
For the scalar sector, we introduce an $SU(2)_L$ doublet $\eta$, singly-charged singlet $\chi^\pm$ 
and doubly-charged singlet $\kappa^{\pm\pm}$ fields.
Among these new fields, only $\kappa^{\pm\pm}$ are assigned to be $Z_2$ even. 
A DM candidate can then be obtained as the lightest neutral $Z_2$ odd
particle; namely, a neutral component of $\eta$. 

New terms in the Lagrangian for lepton Yukawa interactions
and the scalar potential under the $SU(2)_L\times U(1)_Y\times Z_2$ invariance are given by
\begin{align}
-\mathcal{L}_{Y}
&=
+ (y_{\eta})_{i\alpha} \overline{L_L^i} \eta E_R^\alpha  
+ (y_{R})_{\alpha\beta}  \overline{E^{c\alpha}_R} E_R^\beta \kappa^{++}
+(y_{L})_{\alpha\beta} \overline{E^{c\alpha}_L} E_L^\beta \kappa^{++}
 +M_{E^\alpha}\overline{ E_L^\alpha} E_R^\alpha+\rm{h.c.},\\ 
\mathcal{V}
&= 
 M_\Phi^2 |\Phi|^2 + M_\eta^2 |\eta|^2 + M_\chi^2 |\chi^+|^2  +
 M_\kappa^2 |\kappa^{++}|^2\notag\\
&+ 
\left(\mu \Phi \eta \chi^- + \mu_\kappa \chi^- \chi^- \kappa^{++} + {\rm h.c.}\right)
 \nn\\
&
  +\lambda_1 |\Phi|^{4} 
  + \lambda_2 |\eta|^{4} 
  + \lambda_3 |\Phi|^2|\eta|^2 
  + \lambda_4 (\Phi^\dagger \eta)(\eta^\dagger \Phi)
 +\Bigl[\lambda_5(\Phi^\dag\eta)^2+{\rm h.c.}\Bigr]\nn\\
& +\lambda_{\Phi \chi}  |\Phi|^2|\chi^+|^2
  +\lambda_{\Phi\kappa} |\Phi|^2|\kappa^{++}|^2\nn	\\
&  + \lambda_{\eta\chi}  |\eta|^2|\chi^+|^2
  + \lambda_{\eta\kappa}  |\eta|^2|\kappa^{++}|^2
  + \lambda_{\chi\kappa} |\chi^{+}|^2|\kappa^{++}|^2+
 \lambda_{\chi} |\chi^{+}|^4  +
 \lambda_{\kappa} |\kappa^{++}|^4,
\label{HP}
\end{align}
where $y_L^{}$ and $y_R^{}$ are the symmetric $3\times 3$ complex matrices. 
In the scalar potential, $\mu$, $\mu_\kappa$ and $\lambda_5$ can be chosen to be real 
without any loss of generality by renormalizing the phases to scalar bosons. 
Notice here that although the $(\overline{e^{ci}_R} e_R^j \kappa^{++}+\text{h.c.})$
terms are allowed in general\footnote{We can forbid these terms by introducing an additional $U(1)$ symmetry. 
However, such a symmetry also forbids the $\lambda_5$ term in the
potential in Eq.~(\ref{HP}) which is necessary to avoid the constraint
from the direct detection experiment for DM.},  
they do not contribute to neutrino mass generations. 
We thus neglect these terms for simplicity throughout the paper. 

After the electroweak symmetry breaking, the scalar fields can be parameterized as 
\begin{align}
&\Phi =\left[
\begin{array}{c}
G^+\\
\frac1{\sqrt2}(v+h+iG^0)
\end{array}\right],\quad 
\eta =\left[
\begin{array}{c}
\eta^+\\
\frac1{\sqrt2}(\eta_{R}+i\eta_{I})
\end{array}\right].
\label{component}
\end{align}
where $v~\simeq 246$ GeV is the vacuum expectation value (VEV) of the Higgs doublet, and $G^\pm$
and $G^0$ are respectively the Nambu-Goldstone bosons 
which are absorbed by the longitudinal components of $W^\pm$ and $Z$ bosons.  
The other scalar fields are not supposed to have a non-zero VEV, otherwise the electromagnetic interactions
or $Z_2$ symmetry spontaneously breaks down. 

The mass of the Higgs boson $h$ is obtained after taking the vacuum condition; i.e., 
$\partial\mathcal{V}/\partial\Phi|_{\mathrm{VEV}}=0$ just like the SM Higgs boson as 
$m_h^2 = 2\lambda_1 v^2$. 
In our model, only the singly-charged scalar states $\eta^\pm$ and
$\chi^\pm$ can mix each other among the new scalar bosons. 
The mass matrix for the singly-charged scalar states $M_\pm^2$ in the
basis of ($\eta^\pm$, $\chi^\pm$) is given by 
\begin{align}
M_\pm^2 &= \left(%
\begin{array}{cc}
M^2_\eta+\frac{v^2}{2}\lambda_3   & \frac{\mu v}{\sqrt2} \\
 \frac{\mu v}{\sqrt2}  & M^2_\chi + \frac{v^2}{2}\lambda_{\Phi\chi} \\
\end{array}%
\right).
\end{align}
The mass eigenstates $H_1^\pm$ and $H_2^\pm$ are defined by introducing the
mixing angle $\theta$ as 
\begin{align}
\left(\begin{array}{c}
\eta^\pm \\
\chi^\pm 
\end{array}\right)=
\left(\begin{array}{cc}
\cos\theta  & -\sin\theta\\
\sin\theta & \cos\theta 
\end{array}\right)
\left(\begin{array}{c}
H_1^\pm \\
H_2^\pm 
\end{array}\right),
\end{align}
where $\theta$ and the mass eigenvalues are given by 
\begin{align}
m_{H_1^\pm} &= (M_\pm^2)_{11}\cos^2\theta + (M_\pm^2)_{22}\sin^2\theta + (M_\pm^2)_{12}\sin2\theta , \\
m_{H_2^\pm} &= (M_\pm^2)_{11}\sin^2\theta + (M_\pm^2)_{22}\cos^2\theta - (M_\pm^2)_{12}\sin2\theta , \\
\tan 2\theta &=\frac{2(M_\pm^2)_{12}}{(M_\pm^2)_{11}-(M_\pm^2)_{22}}. 
\end{align}
All the other masses of scalar bosons are calculated as 
\begin{align}
m^{2}_{\kappa^{\pm\pm}} &= M_\kappa^{2}  + \frac12 \lambda_{\Phi \kappa} v^{2}, \\ 
m^2_{\eta_{R}} &= 
M_\eta^{2} + \frac12\left( \lambda_3  + \lambda_4 + 2\lambda_5\right)v^2, \\  
m^2_{\eta_{I}} &=
M_\eta^{2} + \frac12\left( \lambda_3  + \lambda_4 - 2\lambda_5\right)v^2. 
\end{align}

\subsection{Dark Matter}

The neutral component of $\eta$ is the unique DM candidate in the
model. Taking $\lambda_5>0$, the DM can be identified as $\eta_I$ which
should satisfy the observed thermal relic density~\cite{Ade:2013lta}. 
The DM mass is limited as $m_{\eta_I}\lesssim80~\mathrm{GeV}$ and
$m_{\eta_I}\gtrsim500~\mathrm{GeV}$ from the constraint of thermal
relic density of DM~\cite{LopezHonorez:2006gr, Hambye:2009pw}. When the DM is in the low
mass region $m_{\eta_I}\lesssim80~\mathrm{GeV}$ (below the $W$ or $Z$ threshold),
the main annihilation process is $\eta_I\eta_I\to h\to f\overline{f}$
where $b\overline{b}$ is dominant in the final state fermions
$f\overline{f}$ due to the size of the Yukawa coupling. 
In addition to the annihilation process, the co-annihilation with
$\eta_R$ mediated by $Z$ boson, $\eta_R\eta_I\to Z\to f\overline{f}$ is
effective if the mass difference among them is less than around $10\%$. 
The flavor of the final state fermions are universal unlike the
annihilation process. 
Typically the DM mass should be around the SM Higgs resonance; i.e., $m_h/2\simeq 63$ GeV, to satisfy
the thermal DM relic density in the low DM mass region. 
On the other hand, the annihilation processes $\eta_I\eta_I\to W^+W^-,\,ZZ$ and $hh$ can be dominant when the DM mass is above the thresholds. 
In the region of the DM mass $80~\mathrm{GeV}\lesssim
m_{\eta_I}\lesssim500~\mathrm{GeV}$, the predicted relic density is
strongly suppressed since the cross section which is fixed by the gauge
coupling is too large. 
We are interested in the low DM mass region because such a low DM mass
is favored from inducing the large muon anomalous magnetic moment as will
be discussed later. Thus the DM mass is fixed to $m_h/2$ in the following discussion. 

There is the constraint from direct detection experiments since the elastic
scattering process with nuclei is induced via Higgs boson exchange. 
The relevant parameters in the model are $\lambda_3$ $\lambda_4$ and
$\lambda_5$, and they are roughly restricted as 
$\lambda_3+\lambda_4-\lambda_5\lesssim0.03$ in the low DM mass region~\cite{LopezHonorez:2006gr}
from the recent LUX experiment~\cite{Akerib:2013tjd}. 
Moreover, magnitude of the mass degeneracy between $\eta_R$ and $\eta_I$ is 
constrained because the inelastic scattering process $\eta_I N\to \eta_R N$
can occur via $Z$ boson exchange if the mass splitting is small enough. 
The mass splitting as small as $100~\mathrm{keV}$ is typically ruled out
from the $Z$ exchange process~\cite{TuckerSmith:2001hy}.

\subsection{$S$, $T$ and $U$ Parameters}

We consider the constraints from the electroweak $S$, $T$ and $U$
parameters proposed by Peskin and Takeuchi~\cite{stu}. 
The new contributions to the $S$, $T$ and $U$ parameters are given by
\begin{align}
S_{\text{new}}&=\frac{1}{4\pi m_Z^2}
\Big[c_\theta^2(c_\theta^2-2)F(m_Z^2;m_{H_1^\pm},m_{H_1^\pm})+s_\theta^2(s_\theta^2-2)F(m_Z^2;m_{H_2^\pm},m_{H_2^\pm})\notag\\
&+2s_\theta^2c_\theta^2[F(m_Z^2;m_{H_1^\pm},m_{H_2^\pm})-F(0;m_{H_1^\pm},m_{H_2^\pm})]
+[F(m_Z^2;m_{\eta_R},m_{\eta_I})-F(0;m_{\eta_R},m_{\eta_I})]\Big]
,\\
T_{\text{new}}
&=
\frac{1}{16\pi^2 \alpha_{{\rm em}} v^2} 
\Big[ 
c^2_\theta F(0;m_{H_1^\pm}, m_{\eta_{R}})+  c^2_\theta F(0;m_{H_1^\pm},m_{\eta_{I}})
+s^2_\theta F(0;m_{H_2^\pm}, m_{\eta_{R}})\notag\\
&\qquad\qquad\qquad
+ s^2_\theta F(0;m_{H_2^\pm}, m_{\eta_{I}})
-F(0;m_{\eta_{I}}, m_{\eta_{R}})-2s^2_\theta c^2_\theta F(0;m_{H_1^\pm}, m_{H_2^\pm}) \Big],\\
U_{\text{new}}&=-\frac{1}{4\pi m_Z^2}
\Bigg\{c_\theta^4F(m_Z^2;m_{H_1^\pm},m_{H_1^\pm})+s_\theta^4F(m_Z^2;m_{H_2^\pm},m_{H_2^\pm})\notag\\
&+2s_\theta^2c_\theta^2[F(m_Z^2;m_{H_1^\pm},m_{H_2^\pm})-F(0;m_{H_1^\pm},m_{H_2^\pm})]
+[F(m_Z^2;m_{\eta_R},m_{\eta_I})-F(0;m_{\eta_R},m_{\eta_I})]\notag\\
&-\frac{m_Z^2}{m_W^2}\Big[c_\theta^2\left(F(m_W^2;m_{H_1^\pm},m_{\eta_R})-F(0;m_{H_1^\pm},m_{\eta_R})
+F(m_W^2;m_{H_1^\pm},m_{\eta_I})-F(0;m_{H_1^\pm},m_{\eta_I})\right)\notag\\
&+s_\theta^2\left(F(m_W^2;m_{H_2^\pm},m_{\eta_R})-F(0;m_{H_2^\pm},m_{\eta_R})
+F(m_W^2;m_{H_2^\pm},m_{\eta_I})-F(0;m_{H_2^\pm},m_{\eta_I})\right)
\Big]\Bigg\},
\end{align}
where $c_\theta\equiv\cos\theta$ and $s_\theta\equiv\sin\theta$, and $\alpha_{\mathrm{em}}\simeq 1/137$ is the electromagnetic fine structure constant.
The function $F(p^2;m_1,m_2)$ is given by 
\begin{align}
F(p^2;m_1,m_2)&=\int_0^1 dx \Big[(2x-1)(m_1^2-m_2^2)+(2x-1)^2p^2\Big]\ln \Big[xm_1^2+(1-x)m_2^2-x(1-x)p^2\Big],
\end{align}
This function is reduced to the simple form in the case of $p^2=0$ as 
\begin{align}
F(0;m_1, m_2)&=
\frac{m_1^2+m_2^2}{2}-\frac{m_1^2m_2^2}{m_1^2-m_2^2}\ln\left(\frac{m_1^2}{m_2^2}\right), 
\end{align}
which gives zero in the case of $m_1 = m_2$. 
The experimental deviations from the SM predictions in the $S$ and $T$ parameters under
$m_h=126$ GeV and $U=0$ are given by~\cite{Baak:2012kk}\footnote{
Typically, the magnitude of $U_{\text{new}}$ is smaller than 0.01 when 
the prediction of $S_{\text{new}}$ and $T_{\text{new}}$ parameters are
inside the 95\% CL allowed region by the data. } 
\be
S_{\mathrm{new}}=0.05\pm 0.09,\quad T_{\mathrm{new}}=0.08\pm 0.07,
\label{STallowed}
\ee
where the correlation factor between $S_{\text{new}}$ and $T_{\text{new}}$ is +0.91. 

\begin{figure}[t]
\begin{center}
\includegraphics[scale=0.5]{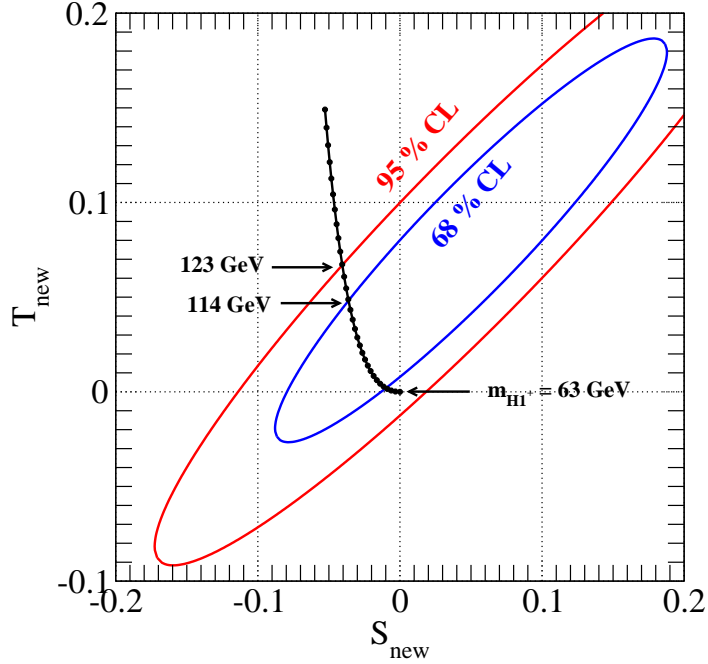}
\end{center}
   \caption{Contour plot for the $S_{\text{new}}$ and $T_{\text{new}}$
 parameters. We take $\theta = 0$ and $m_{\eta_R}=m_{\eta_I}=63$ GeV. 
The blue and red ellipses denote the 68\% and 95\% CL limits, respectively, for the $S$ and $T$ parameters. 
The black dots show the prediction of the $S_{\text{new}}$ and
 $T_{\text{new}}$ parameters, where $m_{H_1^\pm}$ is valid from 63 GeV to
 153 GeV with the 3 GeV interval.}
   \label{st_plot}
\end{figure}

In Fig.~\ref{st_plot}, we show the prediction of the $S_{\text{new}}$ and $T_{\text{new}}$ parameters. 
The 68\% and 95\% CL limits for the $S_{\text{new}}$ and
$T_{\text{new}}$ parameters are respectively denoted by the blue and red
ellipses. 
We take $\theta=0$ and $m_{\eta_R}= m_{\eta_I}=63$ GeV in this plot.
In that case, $S_{\text{new}}$ and $T_{\text{new}}$ are determined by
fixing $m_{H_1^\pm}$ which corresponds to the mass of $\eta^\pm$. 
The prediction on the $S_\text{new}$-$T_\text{new}$ plane changes from
the origin to the left-upper edge along the black curve 
when $m_{H_1^\pm}$ is valid from 63 GeV to 153 GeV. 
From this figure, we obtain the upper limit on $m_{H_1^\pm}$ to be 114
GeV and 123 GeV with the 68\% CL and 95\% CL, respectively. 
The upper limit on $m_{H_1^\pm}$ will be relaxed with several dozens GeV if
$U_{\mathrm{new}}\neq0$ is taken into account~\cite{Baak:2012kk}.

\section{Neutrino Mass and Muon Anomalous Magnetic Moment}

\subsection{Neutrino mass matrix}

\begin{figure}[t]
\begin{center}
\includegraphics[scale=0.9]{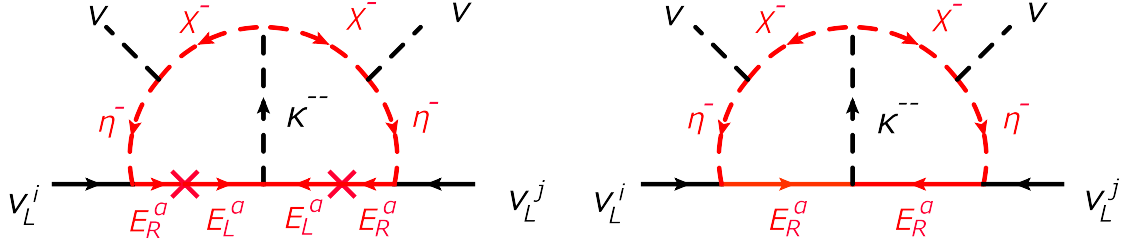}
   \caption{Feynman diagrams for the neutrino mass generation at the two-loop level. 
Particles indicated by the red color have the $Z_2$ odd parity. }
   \label{neutrino-diag}
\end{center}
\end{figure}
The Majorana neutrino mass matrix $m_\nu$ is derived at two-loop level from the
diagrams depicted in Fig.~\ref{neutrino-diag}. The contribution of the
left diagram is given by 
\begin{align}
(m_{\nu}^L)_{ij}
&=
\frac{\mu_\kappa\sin^2 2\theta}{4(16\pi^2)^2} \sum_{\alpha=1}^{3} \sum_{\beta=1}^{3}
\left[
 \frac{(y_\eta)_{i\alpha}M_{E^\alpha}(y_L)^*_{\alpha\beta}(M_{E^\beta})(y_{\eta})_{\beta
 j}}{M_{E^\alpha}^2}\right] 
\nn\\
&\times 
\left[
F_L\left(
\frac{m_{H_1^\pm}^2}{M_{E^\alpha}^2},
\frac{m_{\kappa^{\pm\pm}}^2}{M_{E^\alpha}^2},
\frac{m_{H_1^\pm}^2}{M_{E^\alpha}^2},
\frac{M_{E^\beta}^2}{M_{E^\alpha}^2}
\right)
-
F_L\left(
\frac{m_{H_2^\pm}^2}{M_{E^\alpha}^2},
\frac{m_{\kappa^{\pm\pm}}^2}{M_{E^\alpha}^2},
\frac{m_{H_1^\pm}^2}{M_{E^\alpha}^2},
\frac{M_{E^\beta}^2}{M_{E^\alpha}^2}
\right)\right.\nonumber\\
&\left.\qquad
-F_L\left(
\frac{m_{H_1^\pm}^2}{M_{E^\alpha}^2},
\frac{m_{\kappa^{\pm\pm}}^2}{M_{E^\alpha}^2},
\frac{m_{H_2^\pm}^2}{M_{E^\alpha}^2},
\frac{M_{E^\beta}^2}{M_{E^\alpha}^2}
\right)
+F_L\left(
\frac{m_{H_2^\pm}^2}{M_{E^\alpha}^2},
\frac{m_{\kappa^{\pm\pm}}^2}{M_{E^\alpha}^2},
\frac{m_{H_2^\pm}^2}{M_{E^\alpha}^2},
\frac{M_{E^\beta}^2}{M_{E^\alpha}^2}
\right)
\right],
\end{align}
and that of the right diagram is given by
\begin{align}
(m_\nu^R)_{ij}&=
\frac{\mu_\kappa\sin^2 2\theta}{4(16\pi^2)^2}
\sum_{\alpha=1}^{3} \sum_{\beta=1}^{3}
\left[(y_\eta)_{i\alpha}(y_R)^*_{\alpha\beta}(y_{\eta})_{\beta j} \right]\nonumber\\
&\times\left[ 
-F_R\left(
\frac{m_{H_1^\pm}^2}{M_{E^\alpha}^2},
\frac{m_{\kappa^{\pm\pm}}^2}{M_{E^\alpha}^2},
\frac{m_{H_1^\pm}^2}{M_{E^\alpha}^2},
\frac{M_{E^\beta}^2}{M_{E^\alpha}^2}
\right)
+
F_R\left(
\frac{m_{H_1^\pm}^2}{M_{E^\alpha}^2},
\frac{m_{\kappa^{\pm\pm}}^2}{M_{E^\alpha}^2},
\frac{m_{H_2^\pm}^2}{M_{E^\alpha}^2},
\frac{M_{E^\beta}^2}{M_{E^\alpha}^2}
\right)\right.\nonumber\\
&\left.\qquad
+F_R\left(
\frac{m_{H_2^\pm}^2}{M_{E^\alpha}^2},
\frac{m_{\kappa^{\pm\pm}}^2}{M_{E^\alpha}^2},
\frac{m_{H_1^\pm}^2}{M_{E^\alpha}^2},
\frac{M_{E^\beta}^2}{M_{E^\alpha}^2}
\right)
-F_R\left(
\frac{m_{H_2^\pm}^2}{M_{E^\alpha}^2},
\frac{m_{\kappa^{\pm\pm}}^2}{M_{E^\alpha}^2},
\frac{m_{H_2^\pm}^2}{M_{E^\alpha}^2},
\frac{M_{E^\beta}^2}{M_{E^\alpha}^2}
\right)
\right].
\label{eq:neutrinomass}
\end{align}
The loop functions $F_L$ and $F_R$ are computed as
\begin{align} 
F_L\left(X_1,X_2,X_3,X_4\right) 
&=
\int_0^1 dxdydz\frac{\delta(x+y+z-1)}{z(z-1)}\int_0^1 dx'dy'dz'
\frac{\delta(x'+y'+z'-1)}{x'+y' X_1-z' \Delta(X_2,X_3,X_4)}, \\
F_R\left(X_1,X_2,X_3,X_4 \right) 
&=
2\int_0^1 dxdydz\frac{\delta(x+y+z-1)}{z-1}\int_0^1
 dx'dy'dz'\delta(x'+y'+z'-1)\nonumber\\
&\qquad\times
\ln\left[x'+y'X_1-z'\Delta(X_2,X_3,X_4)\right],
\end{align}
where
\begin{align}
\Delta(X_2,X_3,X_4)=\frac{xX_4+yX_3+zX_2}{z(z-1)}.
\end{align}
The total neutrino mass matrix is given by
$(m_\nu)_{ij}=(m_\nu^L)_{ij}+(m_\nu^R)_{ij}$. 
Notice here that the contribution from the left figure in
Fig.~\ref{neutrino-diag} is reduced to the neutrino
mass matrix given in the original Zee-Babu
model in the limit of $M_{E^\alpha}\to0$~\cite{AristizabalSierra:2006gb}.

When we assume that 
$y_\eta$ is proportional to the unit matrix; i.e., $(y_\eta)_{i\alpha} = y_\eta \times{\bm 1_{3\times 3}}$,
and all the masses for $E^\alpha$ are degenerate; $M_{E^\alpha}\,(\alpha=1,2,3)=M_E$, 
the neutrino mass matrix is simply rewritten as 
\begin{align}
(m_\nu)_{ij} = C_L (y_L)_{ij} - C_R (y_R)_{ij} , 
\end{align}
where 
\begin{align}
&C_{L,R} = \frac{\mu_\kappa }{4(16\pi^2)^2}y_\eta^2\sin^22\theta\left[
F_{L,R}\left(x_{H_1^\pm},  x_{\kappa^{\pm\pm}},x_{H_1^\pm},1 \right)
-
F_{L,R}\left(x_{H_2^\pm},  x_{\kappa^{\pm\pm}},x_{H_1^\pm},1 \right)\right.\nonumber\\
&\left.\qquad
-F_{L,R}\left(x_{H_1^\pm},  x_{\kappa^{\pm\pm}},x_{H_2^\pm},1 \right)
+F_{L,R}\left(x_{H_2^\pm},  x_{\kappa^{\pm\pm}},x_{H_2^\pm},1 \right)
\right], ~ \text{with} ~  x_\varphi=m_\varphi^2/M_{E}^2.
\end{align}

The neutrino mass matrix is diagonalized by introducing the
Pontecorvo-Maki-Nakagawa-Sakata matrix $U_{\text{PMNS}}$ as~\cite{pmns} 
\begin{align}
U_{\text{PMNS}}^T \, m_\nu \, U_{\text{PMNS}} = \text{diag}(m_1,m_2,m_3). \label{mns}
\end{align}
By using the following 
best fit values of the neutrino mixing angles and squared mass
differences assuming the normal (inverted) mass
hierarchy~\cite{neutrino_mixing}; 
\begin{align}
&\sin^2\theta_{12} = 0.323\,(0.323),~
\sin^2\theta_{23} = 0.567\,(0.573),~
\sin^2\theta_{13} = 0.0234\,(0.0240),~\\
& \Delta m_{21}^2 = 7.60\,(7.60) \times 10^{-5} \text{ eV}^2,~
|\Delta m_{31}|^2=2.48\,(2.38) \times 10^{-3}\text{ eV}^2, 
\end{align}
we obtain the $U_{\text{PMNS}}$ matrix elements as 
\begin{align}
U_{\text{PMNS}} = \left(\begin{array}{ccc}
0.813\,(0.813) & 0.562\,(0.561) & 0.153\,(0.155) \\
-0.469\,(-0.468)& 0.476\,(0.471) & 0.744\,(0.748) \\ 
0.345\,(0.347) & -0.677\,(-0.680)& 0.650\,(0.646)
\end{array}\right), \label{mns_exp}
\end{align}
where we neglect the CP-violating phase, which is experimentally allowed
within the 2-$\sigma$ level~\cite{neutrino_mixing}. 
In addition, using the central value of the sum of neutrino masses 0.36 eV~\cite{neutrino_sum}, 
the eigenvalues for the neutrino masses are determined in the normal (inverted) hierarchy as 
\begin{align}
m_1=0.116\,(0.124)\text{ eV},~ m_2= 0.117\,(0.123)\text{ eV},~ m_3 =
 0.127\,(0.113)\text{ eV}. 
\label{mass_eigen} 
\end{align}

From Eqs.~(\ref{mns}), (\ref{mns_exp}) and (\ref{mass_eigen}), and taking $y_L^{}=y_R^{}~(=\bar{y})$, 
we obtain the matrix elements of $\bar{y}$ in the normal (inverted) hierarchy;  
\begin{align}
\bar{y}_{ij} &= \frac{m_{\text{max}}}{C_L-C_R}U_{\text{PMNS}}\,\text{diag}(m_1/m_{\text{max}},m_2/m_{\text{max}},m_3/m_{\text{max}})\,U_{\text{PMNS}}^T\notag\\
&=  \frac{m_{\text{max}}}{C_L-C_R} 
\left(\begin{array}{ccc}
0.922\,(0.997)& 0.00985\,(-0.0104) & 0.00703\,(-0.00744)\\ 
0.00985\,(-0.0104)& 0.965\,(0.952) & 0.0381\,(-0.0397)\\
0.00703\,(-0.00744)& 0.0381\,(-0.0397) & 0.955\,(0.964)
\end{array}\right), \label{mixing}
\end{align}
where $m_{\text{max}}$ is the largest eigenvalue of neutrino masses;
i.e., $m_{\text{max}}=m_3$ ($m_1$). 
The order of magnitude of the parameter $|C_L-C_R|$ is estimated as 
\begin{align}
|C_L-C_R| = 0.1\text{ eV}\times \biggl(\frac{y_\eta}{1.0}\biggr)^2
 \left(\frac{|\mu_\kappa|}{1 \text{ TeV}}\right)
 \left(\frac{\sin2\theta}{10^{-4}}\right)^2, 
\label{clr}
\end{align}
where typical magnitudes of the loop functions $F_{L,R}$ are fixed to be $\mathcal{O}(1)$.

\subsection{Muon anomalous magnetic moment}

\begin{figure}[t]
\begin{center}
\includegraphics[scale=0.7]{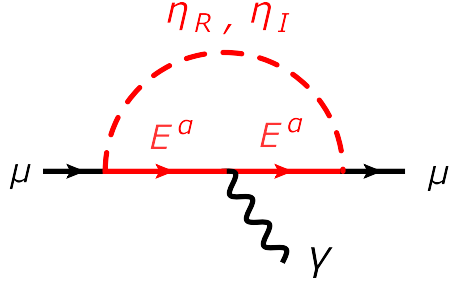}
   \caption{Feynman diagram for the new contribution to the muon $g-2$. }
   \label{g2-diag}
\end{center}
\end{figure}

The muon anomalous magnetic moment (muon $g-2$) has been 
measured at Brookhaven National Laboratory. 
The current average of the experimental results is given by~\cite{bennett}
\begin{align}
a^{\rm exp}_{\mu}=11 659 208.0(6.3)\times 10^{-10}. \notag
\end{align}
It has been well known that there is a discrepancy between the
experimental data and the prediction in the SM. 
The difference $\Delta a_{\mu}\equiv a^{\rm exp}_{\mu}-a^{\rm SM}_{\mu}$
was calculated in Ref.~\cite{discrepancy1} as 
\begin{align}
\Delta a_{\mu}=(29.0 \pm 9.0)\times 10^{-10}, \label{dev1}
\end{align}
and it was also derived in Ref.~\cite{discrepancy2} as
\begin{align}
\Delta a_{\mu}=(33.5 \pm 8.2)\times 10^{-10}. \label{dev2}
\end{align}
The above results given in Eqs. (\ref{dev1}) and (\ref{dev2}) correspond
to $3.2\sigma$ and $4.1\sigma$ deviations, respectively. 

In our model, there are new contributions to $\Delta a_\mu$ as shown in
Fig.~\ref{g2-diag}. 
These contributions are calculated as
\begin{align}
\Delta a_\mu &= \frac{1}{32\pi^2}\sum_{\alpha=1}^3
|y_{\eta}^{2\alpha}|^2\left(\frac{m_\mu}{M_{E^\alpha}}\right)^2\left[
G\left(\frac{m_{\eta_R}^2}{M_{E^\alpha}^2}\right)
+G\left(\frac{m_{\eta_I}^2}{M_{E^\alpha}^2}\right)\right], 
\label{eq:muon-g-2}
\end{align}
where
\begin{align}
G(x)&= \frac{1-6x+3x^2+2x^3-6x^2\ln x}{6(1-x)^4}.
\end{align}
Under the same assumption taken in the previous subsection, 
we obtain
\begin{align}
\Delta a_\mu &= \frac{y_\eta^2}{16\pi^2}
\left(\frac{m_\mu}{M_E}\right)^2G\left(\frac{m_{\eta^0}^2}{M_{E}^2}\right), \label{amu2}
\end{align}
where we neglect the mass difference between $\eta_R$ and $\eta_I$;
i.e., $m_{\eta_R}=m_{\eta_I}\,\equiv m_{\eta^0}$. 

\begin{figure}[t]
\begin{center}
\includegraphics[scale=0.45]{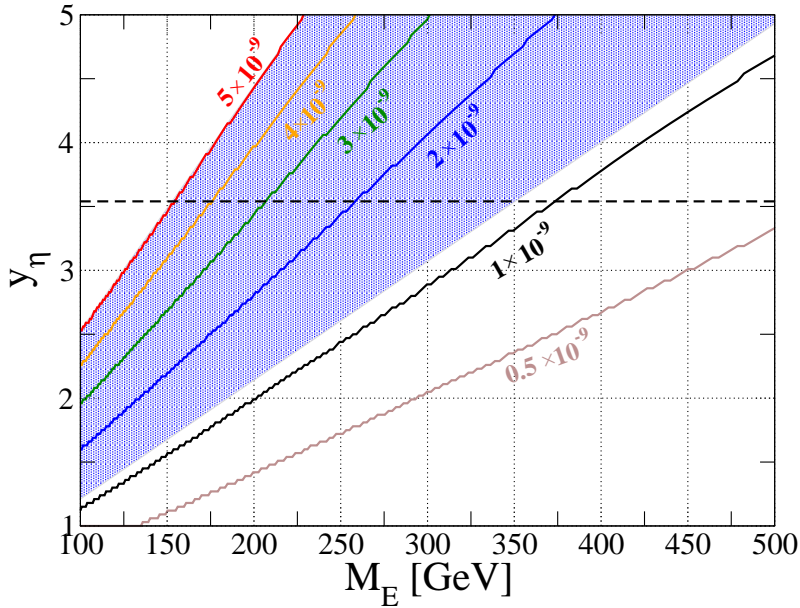}
\caption{Contour plots for $\Delta a_\mu$ using Eq.~(\ref{amu2}) on the
 $M_E$-$y_\eta$ plane in the case of $m_{\eta^0}=63$ GeV. 
In the shaded regions, $1.1 \times 10^{-9}<\Delta a_\mu < 5.0\times
 10^{-9}$ is satisfied, where the lower and upper limits on $\Delta a_\mu$ are respectively derived from
 Eqs.~(\ref{dev1}) and (\ref{dev2}) with taking into account the
 2-$\sigma$ error. The horizontal dashed line denotes the upper limit for
 $y_\eta$ from the requirement of perturbativity; i.e., $y_\eta^2/(4\pi)\leq 1$. 
}
   \label{g2-fig}
\end{center}
\end{figure}

In Fig.~\ref{g2-fig}, the prediction of $\Delta a_\mu$ is shown on the
$M_E$-$y_\eta$ plane in the case of $m_{\eta^0}=63$ GeV. 
From Eqs.~(\ref{dev1}) and (\ref{dev2}), we can obtain the lower and
upper limits on $\Delta a_\mu$ as $1.1 \times 10^{-9}<\Delta a_\mu <
5.0\times 10^{-9}$ by allowing up to the $2$-$\sigma$ error. 
In the blue shaded regions, the prediction is inside the above limit. 
Furthermore, if we require perturbativity of the coupling constant $y_\eta$; i.e., $y_\eta^2/(4\pi)\leq 1$, 
the upper limit for $y_\eta$ can be set, which is denoted by the horizontal dashed line.  
When we take the maximal allowed value of $y_\eta$ from perturbativity, 
the region of  $150~\text{GeV} \lesssim M_E \lesssim 350$ GeV is favored by $\Delta a_\mu$. 
In the next section, we discuss the constraint on $M_E$ from the current LHC data. 

In the end of this section, we comment on new contributions to the
lepton flavor violating (LFV) processes such as $\mu\to e\gamma$ in our
model.  
In general, we can consider diagrams which contribute to the LFV processes by replacing the external muons shown in Fig.~\ref{g2-diag} 
with charged-leptons with different flavors with each other. 
Such a contribution is proportional to the off-diagonal element of the $y_\eta$ coupling, so that 
we can avoid constraints from the LFV data as long as we take the diagonal structure of $y_\eta$. 
 
\section{Collider Phenomenology}

In this section, we discuss the collider phenomenology of our model. 
First of all, we specify the mass spectrum according to the previous sections. 
From the physics of DM, we set the mass of the lightest neutral $Z_2$ odd scalar boson $\eta_I$ to be 63 GeV. 
The mass of $\eta_R$ must be larger than that of $\eta_I$ at least order of 100 keV from the direct detection experiments for DM. 
In addition in order to reproduce the collect order of neutrino masses, the mixing
angle $\theta$ has to be as small as $\mathcal{O}(10^{-4})$ as seen
Eq.~(\ref{clr}).  
Such a small mass difference between $\eta_I$ and $\eta_R$ and small angle $\theta$ can be neglected in the collider phenomenology.  
We thus take $m_{\eta_R}=m_{\eta_I}$ and $\theta=0$ for simplicity.  
In that case, from the $S$ and $T$ parameters, the upper limit for the
mass of $\eta^\pm $~($=m_{H_1^\pm}$) is given to be 123 GeV with the 95 \% CL (see
Fig.~\ref{st_plot}). 

\begin{figure}[t]
\begin{center}
\includegraphics[scale=0.45]{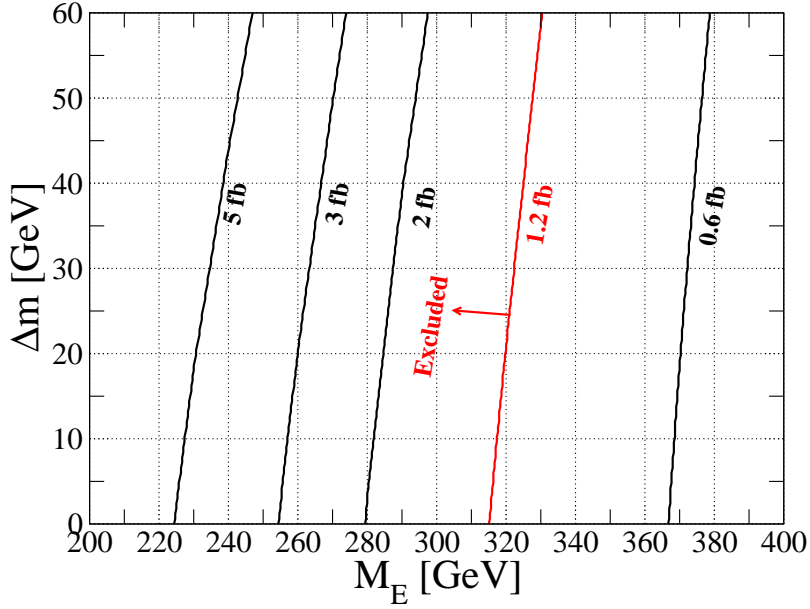}
\caption{Contour plots for the cross section of the $pp\to E_\alpha^+E_\alpha^-\to \ell_\alpha^+\ell_\alpha^-\eta^0\eta^0$ process for each lepton flavour $\alpha$
on the $M_E$-$\Delta m~(\equiv m_{\eta^\pm}-m_{\eta^0})$ 
with the collision energy to be 8 TeV. 
We fix $m_{\eta^0}$ to be 63 GeV. The red contour denotes the cross section of 1.2 fb which corresponds to the upper limit from the slepton search performed in 
Ref.~\cite{slepton}. 
}
   \label{const}
\end{center}
\end{figure}

The mass of vector-like charged leptons $E^\alpha$ can be constrained
from slepton searches at the LHC in the following way. 
In Ref.~\cite{slepton}, the bound on the slepton masses has been given
from the search for $pp\to\tilde{\ell}^+ \tilde{\ell}^-\to \ell^+\ell^-
\tilde{\chi}^0\tilde{\chi}^0$ process, where $\tilde{\ell}^\pm$ and
$\tilde{\chi}^0$ are 
the charged slepton and the neutralino, respectively, and $\ell$ is $e$ or $\mu$. 
For the case where produced sleptons are purely left-handed, the lower
limit on the mass of $\tilde{\ell}$ has been given to be about 300 GeV from the data with the integrated luminosity to be 20.3 fb$^{-1}$ and the collision
energy to be 8 TeV . 
In our model, there are two decay modes of $E^\alpha$; i.e., $E^\alpha \to \nu^\alpha \eta^\pm$
and  $E^\alpha \to \ell^{\alpha \pm} \eta^0$ ($\eta^0$ is $\eta_R$ or $\eta_I$) via the Yukawa coupling $y_\eta$
which is assumed to be proportional to the unit matrix
$(y_\eta)_{i\alpha}=y_\eta\times {\bm 1}_{3\times 3}$. 
Therefore, the similar final state as in the slepton pair production is obtained by 
$pp \to E^{\alpha +}E^{\alpha -}\to \ell^+\ell^- \eta^0\eta^0$. 
The cross section of this process can be estimated as
\begin{align}
\sigma(pp\to  \ell^+\ell^- \eta^0\eta^0) = \sigma(pp\to  E^{\alpha +}E^{\alpha -})\times \mathcal{B}(E^{\alpha \pm}\to \ell^{\alpha\pm}\eta^0 )^2, 
\end{align}
where $\sigma(pp\to  E^{\alpha +}E^{\alpha -})$ is the pair production cross section of $E^{\alpha}$, and 
$\mathcal{B}(E^{\alpha \pm}\to \ell^{\alpha\pm}\eta^0 )$ is the
branching fraction for the $E^{\alpha \pm}\to \ell^{\alpha\pm}\eta^0$
mode. 
When the mass of left-handed slepton is taken to be 300 GeV, the pair
production cross section is calculated to be about 1.2 fb for each flavour
of $\tilde{\ell}$, where we use 
{\tt CalcHEP}~\cite{calchep} and {\tt CTEQ6L} for the parton distribution function. 
We thus can set the lower limit on the mass of $E^\alpha$ by requiring that 
the cross section $\sigma(pp\to  \ell^+\ell^- \eta^0\eta^0)$ does not exceed 1.2 fb.

In Fig.~\ref{const}, we show the contour plots of the cross section of
the process $pp\to E_\alpha^+E_\alpha^-\to \ell_\alpha^+\ell_\alpha^-\eta^0\eta^0$ for each lepton flavour $\alpha$ on the 
$M_E$-$\Delta m~(\equiv m_{\eta^\pm}-m_{\eta^0})$ plane. 
Again, the cross section is calculated by using {\tt CalcHEP} and {\tt CTEQ6L}. 
The upper limit on the cross section from the slepton search is shown as
the red curve, so that the left region from the red curve is excluded. 
We can see that the cross section slightly increases as $\Delta m$ is
getting a large value, because the branching fraction of 
$E_\alpha^\pm \to \ell_\alpha^\pm \eta^0$ gets a small enhancement. 
By looking at the red curve, we find that the lower bound on $M_E$ is given as about 315 GeV to 330 GeV depending on 
the value of $\Delta m$. 
Therefore, there are allowed regions by the LHC bound, where 
the discrepancy in the muon $g-2$ can be explained.

\begin{figure}[t]
\begin{center}
\includegraphics[scale=0.7]{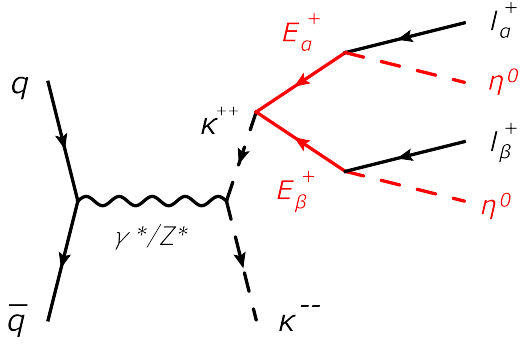}
\end{center}
\caption{Feynman diagram for the signal process in the model. The decay of $\kappa^{--}$ can be 
the same as that of  $\kappa^{++}$ by replacing $\ell_\alpha^+$ with $\ell_\alpha^-$. }
\label{signal}
\end{figure}

In the following discussion, we take $M_E=315$ GeV and $\Delta m\simeq 0$ \footnote{If we exactly take $\Delta m=0$, then 
$\eta^\pm$ cannot decay into the other particles. 
If there is the non-zero mass difference, $\eta^\pm$ can decay into $W^{\pm *}\eta^0$.}. 
In addition, we assume that 
the mass of $\kappa^{\pm\pm}$ is larger than
$2M_E$ and $\chi^\pm$ are heavier than $\kappa^{\pm\pm}$. 
In that case, we expect that the event with the same-sign dilepton plus
missing transverse momentum appears from the 
pair production of $\kappa^{\pm\pm}$ as shown in Fig.~\ref{signal}. 
The cross section of this process is calculated by 
\begin{align}
\sigma(pp\to \ell_\alpha^+\ell_\beta^+E_T\hspace{-4.3mm}/\hspace{3mm}
 \kappa^{--})& = \sigma(pp\to \kappa^{++}\kappa^{--})\times
 \mathcal{B}(\kappa^{++}\to E_\alpha^+ E_\beta^+)\times
\mathcal{B}(E^\pm \to \ell^\pm\eta^0)^2\notag\\
&\simeq \frac{1}{4}\sigma(pp\to \kappa^{++}\kappa^{--})\times
 \mathcal{B}(\kappa^{++}\to E_\alpha^+ E_\beta^+). 
\label{sig-cross}
\end{align}
where $\sigma(pp\to \kappa^{++}\kappa^{--})$ is the pair production cross section of $\kappa^{\pm\pm}$, 
and $\mathcal{B}(\kappa^{++}\to E_\alpha^+ E_\beta^+)$ is the branching fraction of the $\kappa^{++}\to E_\alpha^+ E_\beta^+$ mode. 
Under the assumption of $y_L=y_R~(=\bar{y})$ which was taken in the previous section, 
the decay rate of $\kappa^{++}\to E_\alpha^+ E_\beta^+$ mode is calculated by 
\begin{align}
\Gamma(\kappa^{++}\to E_\alpha^+ E_\beta^+)= S_{\alpha\beta}\frac{|\bar{y}_{\alpha\beta}|^2}{4\pi}m_{\kappa^{\pm\pm}}\left(1-\frac{4M_E^2}{m_{\kappa^{\pm\pm}}^2}\right)^{3/2}, 
\end{align}
where $S_{\alpha\beta}=2~(1)$ for $\alpha \neq \beta$ ($\alpha=\beta$). 
From the above formula, the branching fraction is simply determined by $\bar{y}_{\alpha\beta}$ as 
\begin{align}
\mathcal{B}(\kappa^{++}\to E_\alpha^+ E_\beta^+)= \frac{S_{\alpha\beta}\,|\bar{y}_{\alpha\beta}|^2}{|\bar{y}_{ee}|^2
+|\bar{y}_{\mu\mu}|^2+|\bar{y}_{\tau\tau}|^2+2(|\bar{y}_{e\mu}|^2+|\bar{y}_{e\tau}|^2+|\bar{y}_{\mu\tau}|^2)}. 
\end{align}
Each matrix element of $\bar{y}_{\alpha\beta}$ is given by using the neutrino mixing data as expressed in Eq.~(\ref{mixing}).
In Table~\ref{tab:2}, the branching fraction of $\kappa^{++}$ is listed. 
Because the Yukawa coupling $\bar{y}$ determined from Eq.~(\ref{mixing}) is almost the unite matrix, 
the decay of $\kappa^{++}$ into the same flavour of $E_\alpha$ is dominant. 
This results in the final state of the signal process shown in Fig.~\ref{signal} with the same-sign dilepton with the same-flavour 
and missing transverse energy, and each of three lepton flavours in the final state appears almost the same probability with each other. 
With the 14 TeV energy at the LHC, 
the cross section for the pair production ($pp\to\kappa^{++}\kappa^{--}$) 
and that for the process expressed in Eq.~(\ref{sig-cross}) for each lepton flavour of the final state are also shown in Table~\ref{tab:2}.
We can see that about 10 events can be obtained for the same-sign electron or muon and missing transverse energy final states assuming 300~fb$^{-1}$ 
for the integrated luminosity. 

\begin{center}
\begin{table}[t]
\begin{tabular}{l||c|c|c|c|c|c}\hline\hline   
Mode ($\alpha,\beta$)&($e,e$)&($\mu,\mu$)&($\tau,\tau$)&($e,\mu$)&($e,\tau$)&($\mu,\tau$) \\\hline 
$\mathcal{B}(\kappa^{++}\to E_\alpha^+ E_\beta^+)$ [\%]&31.5&34.5&33.8&7.20$\times 10^{-3}$&3.67$\times 10^{-3}$&0.108  \\
&(35.1)&(32.0)&(32.8)&(7.61$\times 10^{-3})$&(3.91$\times 10^{-3})$&(0.111)  \\\hline\hline
$\sigma(pp\to \kappa^{++} \kappa^{--})$  [fb]&\multicolumn{6}{c}{0.202~$(m_{\kappa^{\pm\pm}}=650$~GeV)}  \\\cline{2-7}
$\sigma(pp\to \ell_\alpha^+\ell_\beta^+E_T\hspace{-4.3mm}/\hspace{3mm} \kappa^{--})/10^{2}$ [fb]&1.59\,(1.77)&1.74\,(1.62)&1.71\,(1.66)&$\simeq$\,0\,(0)&$\simeq$\,0\,(0)&$\simeq$\,0\,(0)
  \\\hline
$\sigma(pp\to \kappa^{++} \kappa^{--})$  [fb]&\multicolumn{6}{c}{0.0690~$(m_{\kappa^{\pm\pm}}=800$~GeV)}  \\\cline{2-7}
$\sigma(pp\to \ell_\alpha^+\ell_\beta^+E_T\hspace{-4.3mm}/\hspace{3mm} \kappa^{--})/10^{2}$ [fb]&
0.543\,(0.615)&0.595\,(0.552)&0.583\,(0.566)&$\simeq$\,0\,(0)&$\simeq$\,0\,(0)&$\simeq$\,0\,(0)
  \\\hline
$\sigma(pp\to \kappa^{++} \kappa^{--})$ [fb]&\multicolumn{6}{c}{0.0193~$(m_{\kappa^{\pm\pm}}=1000$~GeV)}  \\\cline{2-7}
$\sigma(pp\to \ell_\alpha^+\ell_\beta^+E_T\hspace{-4.3mm}/\hspace{3mm} \kappa^{--})/10^{2}$ [fb]&
0.152\,(0.169)&0.166\,(0.154)&0.163\,(0.158)&$\simeq$\,0\,(0)&$\simeq$\,0\,(0)&$\simeq$\,0\,(0)
  \\\hline\hline
\end{tabular}
\caption{Branching fraction of $\kappa^{++}$,  
the cross section for the pair production of $\kappa^{\pm\pm}$,  
and that for the process expressed in Eq.~(\ref{sig-cross}) are shown  with the collision energy to be 14 TeV. 
The numbers without (with) bracket are the results by using the neutrino mixing data assuming the normal (inverted) hierarchy given in Eq.~(\ref{mixing}). 
}
\label{tab:2}
\end{table}
\end{center}

Finally, we comment on signals from doubly-charged scalar bosons from the other models. 
In the original Zee-Babu model, there are isospin singlet doubly-charged scalar bosons just like $\kappa^{\pm\pm}$ in our model. 
They can directly decay into the same-sign dilepton without missing energies. 
Thus, we expect that the sharp peak appears at the mass of doubly-charged scalar bosons 
in the invariant mass distribution for the same-sign dilepton system~\cite{Babu:2002uu, Nebot:2007bc}. 
Similar signal with the same-sign dilepton can also appear in the HTM~\cite{Cheng:1980qt, Schechter:1980gr,
Lazarides:1980nt, Mohapatra:1980yp, Magg:1980ut}, 
in which there are doubly-charged scalar components in the isospin triplet Higgs field. 
On the other hand, the doubly-charged scalar bosons $\kappa^{\pm\pm}$ in our model do not directly decay into the same-sign dilepton as seen in Fig.~\ref{signal}, 
whose decay products include missing energies due to DM. 
The peak in the invariant mass distribution cannot then be seen in that case, 
and the signal from the decay of $\kappa^{\pm\pm}$ is different from that from doubly-charged scalar bosons in the Zee-Babu model and in the HTM. 
In the HTM, the doubly-charged scalar bosons can also decay into the same-sign W bosons 
when the VEV of triplet Higgs field is taken to be larger than about 0.1-1 MeV~\cite{Perez:2008ha, Chiang:2012dk, Kanemura:2013vxa, kang:2014jia, Kanemura:2014goa}. 
Such a decay mode provides a final state with the same-sign dilepton plus missing energies due to the leptonic decay of the W bosons. 
However, the combination of flavour for the same-sign dilepton equally appears due to the universality of leptonic decay of the weak boson.  
In our model, the same-sign dilepton with the different flavour in the final state is strongly suppressed
due to the structure of neutrino mass matrix. 
Therefore, by measuring the lepton flavour of the same-sign dilepton
event, our model can be distinguished from the HTM. 
One should note that if the interaction we have neglected
$\kappa^{++}\overline{e^c_i}e_j$ in the paper exists, eventually the
same invariant mass distribution with the other models may be obtained.
Thus the assumption that there is no Yukawa coupling
$\kappa^{++}\overline{e^c_i}e_j$ is important to see the difference as discussed above. 

\section{Conclusions}

We have constructed the two-loop induced Zee-Babu type neutrino mass model,
in which a DM candidate is included. 
In our model, the DM is the lighter one of the neutral components of the inert doublet scalar.
The DM mass should be around 63 GeV from the view points of
the thermal relic density and explanation for the discrepancy in the muon $g-2$.
The discrepancy between the experimental value of the muon $g-2$ and its prediction in the SM can be explained due to 
the vector-like charged leptons with the mass of less than about 350 GeV. 
By taking into account the current smuon search at the LHC, 
the lower bound on the mass of vector-like leptons has been taken to be about 315 GeV. 
Therefore, the parameter regions favored by the muon $g-2$ are still allowed by the current LHC data.  
We have discussed the collider phenomenology, especially focusing on the doubly-charged scalar bosons $\kappa^{\pm\pm}$. 
We have found that the main decay mode of $\kappa^{\pm\pm}$
can be the same-sign dilepton with the same-flavour plus missing transverse momentum due to the
structure of neutrino mass matrix. It suggests that measuring an excess of the same-sign dilepton events with the same-flavour, 
we can distinguish our model from the other models which include doubly-charged scalar bosons such as the
original Zee-Babu model and the HTM.

\vspace{0.5cm}
\section*{Acknowledgments}
\vspace{0.5cm}
Author thanks to Prof. Seungwon Baek for fruitful discussions. 
T.~T. acknowledges support from the European ITN project (FP7-PEOPLE-
2011-ITN, PITN-GA-2011-289442-INVISIBLES).
K.~Y. was supported in part by the National Science Council of R.O.C. under Grant No. NSC-101-2811-M-008-014.


\begin{thebibliography}{99}

\bibitem{Higgs_ATLAS} 
  G.~Aad {\it et al.}  [ATLAS Collaboration],
  Phys.\ Lett.\ B {\bf 716}, 1 (2012)
  [arXiv:1207.7214 [hep-ex]]. 

\bibitem{Higgs_CMS} 
  S.~Chatrchyan {\it et al.}  [CMS Collaboration],
  Phys.\ Lett.\ B {\bf 716}, 30 (2012)
  [arXiv:1207.7235 [hep-ex]].

\bibitem{Zee}
 A.~Zee,
 Phys.\ Lett.\  B {\bf 93}, 389 (1980) 
 [Erratum-ibid.\  B {\bf 95}, 461  (1980) ].

\bibitem{Zee-Babu}
 A.~Zee,
 Nucl.\ Phys.\ B {\bf 264}, 99  (1986); 
 K.~S.~Babu,
 Phys.\ Lett.\ B {\bf 203}, 132  (1988). 

\bibitem{Krauss:2002px}
  L.~M.~Krauss, S.~Nasri and M.~Trodden,
  Phys.\ Rev.\  D {\bf 67}, 085002 (2003)
  [arXiv:hep-ph/0210389].

\bibitem{Ma:2006km} 
  E.~Ma,
  Phys.\ Rev.\ D {\bf 73}, 077301 (2006)
  [hep-ph/0601225].

\bibitem{Gu:2007ug} 
  P.~-H.~Gu and U.~Sarkar,
  Phys.\ Rev.\ D {\bf 77}, 105031 (2008)
  [arXiv:0712.2933 [hep-ph]].

\bibitem{Gu:2008zf} 
  P.~-H.~Gu and U.~Sarkar,
  Phys.\ Rev.\ D {\bf 78}, 073012 (2008)
  [arXiv:0807.0270 [hep-ph]].

\bibitem{Aoki:2008av}
  M.~Aoki, S.~Kanemura and O.~Seto,
  Phys.\ Rev.\ Lett.\  {\bf 102}, 051805 (2009)
  [arXiv:0807.0361].

\bibitem{Aoki:2010ib} 
  M.~Aoki, S.~Kanemura, T.~Shindou and K.~Yagyu,
  JHEP {\bf 1007}, 084 (2010)
  [Erratum-ibid.\  {\bf 1011}, 049 (2010)]
  [arXiv:1005.5159 [hep-ph]].
	
\bibitem{Kanemura:2011vm} 
  S.~Kanemura, O.~Seto and T.~Shimomura,
  Phys.\ Rev.\ D {\bf 84}, 016004 (2011)
  [arXiv:1101.5713 [hep-ph]].

\bibitem{Lindner:2011it} 
  M.~Lindner, D.~Schmidt and T.~Schwetz,
  Phys.\ Lett.\ B {\bf 705}, 324 (2011)
  [arXiv:1105.4626 [hep-ph]].

\bibitem{Kanemura:2011jj}
 S.~Kanemura, T.~Nabeshima and H.~Sugiyama,
 Phys.\ Lett.\ B {\bf 703}, 66 (2011)
 [arXiv:1106.2480 [hep-ph]].

\bibitem{Aoki:2011he} 
  M.~Aoki, J.~Kubo, T.~Okawa and H.~Takano,
  Phys.\ Lett.\ B {\bf 707}, 107 (2012)
  [arXiv:1110.5403 [hep-ph]].

\bibitem{Kanemura:2011mw} 
  S.~Kanemura, T.~Nabeshima and H.~Sugiyama,
  Phys.\ Rev.\ D {\bf 85}, 033004 (2012)
  [arXiv:1111.0599 [hep-ph]].


\bibitem{Schmidt:2012yg} 
  D.~Schmidt, T.~Schwetz and T.~Toma,
  Phys.\ Rev.\ D {\bf 85}, 073009 (2012)
  [arXiv:1201.0906 [hep-ph]].

\bibitem{Kanemura:2012rj} 
  S.~Kanemura and H.~Sugiyama,
  Phys.\ Rev.\ D {\bf 86}, 073006 (2012)
  [arXiv:1202.5231 [hep-ph]].

\bibitem{Farzan:2012sa} 
  Y.~Farzan and E.~Ma,
  Phys.\ Rev.\ D {\bf 86}, 033007 (2012)
  [arXiv:1204.4890 [hep-ph]].

\bibitem{Bonnet:2012kz} 
  F.~Bonnet, M.~Hirsch, T.~Ota and W.~Winter,
  JHEP {\bf 1207}, 153 (2012)
  [arXiv:1204.5862 [hep-ph]].

\bibitem{Kumericki:2012bf} 
  K.~Kumericki, I.~Picek and B.~Radovcic,
  JHEP {\bf 1207}, 039 (2012)
  [arXiv:1204.6597 [hep-ph]].

\bibitem{Kumericki:2012bh} 
  K.~Kumericki, I.~Picek and B.~Radovcic,
  Phys.\ Rev.\ D {\bf 86}, 013006 (2012)
  [arXiv:1204.6599 [hep-ph]].

\bibitem{Bouchand:2012dx} 
  R.~Bouchand and A.~Merle,
  JHEP {\bf 1207}, 084 (2012)
  [arXiv:1205.0008 [hep-ph]].

\bibitem{Ma:2012if} 
  E.~Ma,
  Phys.\ Lett.\ B {\bf 717}, 235 (2012)
  [arXiv:1206.1812 [hep-ph]].

\bibitem{Gil:2012ya} 
  G.~Gil, P.~Chankowski and M.~Krawczyk,
  Phys.\ Lett.\ B {\bf 717}, 396 (2012)
  [arXiv:1207.0084 [hep-ph]].

\bibitem{Okada:2012np} 
  H.~Okada and T.~Toma,
  Phys.\ Rev.\ D {\bf 86}, 033011 (2012)
  arXiv:1207.0864 [hep-ph].

\bibitem{Hehn:2012kz} 
  D.~Hehn and A.~Ibarra,
  Phys.\ Lett.\ B {\bf 718}, 988 (2013)
  [arXiv:1208.3162 [hep-ph]].

\bibitem{Dev:2012sg} 
  P.~S.~B.~Dev and A.~Pilaftsis,
  Phys.\ Rev.\ D {\bf 86}, 113001 (2012)
  [arXiv:1209.4051 [hep-ph]].

\bibitem{Kajiyama:2012xg} 
  Y.~Kajiyama, H.~Okada and T.~Toma,
  Eur.\ Phys.\ J.\ C {\bf 73}, 2381 (2013)
  [arXiv:1210.2305 [hep-ph]].

\bibitem{Okada:2012sp} 
  H.~Okada,
  arXiv:1212.0492 [hep-ph].

\bibitem{Gustafsson} 
M.~Gustafsson, J.~M.~No and M.~A.~Rivera,
  Phys.\ Rev.\ Lett.\  {\bf 110}, 211802 (2013)
arXiv:1212.4806 [hep-ph].

\bibitem{Aoki:2013gzs} 
  M.~Aoki, J.~Kubo and H.~Takano,
  Phys.\ Rev.\ D {\bf 87}, no. 11, 116001 (2013)
  [arXiv:1302.3936 [hep-ph]].

\bibitem{Kajiyama:2013zla} 
  Y.~Kajiyama, H.~Okada and K.~Yagyu,
  Nucl.\ Phys.\ B {\bf 874}, 198 (2013)
  [arXiv:1303.3463 [hep-ph]].

\bibitem{Kajiyama:2013rla} 
  Y.~Kajiyama, H.~Okada and T.~Toma,
  Phys.\ Rev.\ D {\bf 88}, 015029 (2013)
  [arXiv:1303.7356].

\bibitem{Kanemura:2013qva} 
  S.~Kanemura, T.~Matsui and H.~Sugiyama,
  Phys.\ Lett.\ B {\bf 727}, 151 (2013)
  [arXiv:1305.4521 [hep-ph]].


\bibitem{Hernandez:2013dta} 
  A.~E.~Carcamo Hernandez, I.~d.~M.~Varzielas, S.~G.~Kovalenko, H.~P\"{a}s and I.~Schmidt,
  Phys.\ Rev.\ D {\bf 88}, 076014 (2013)
  [arXiv:1307.6499 [hep-ph]].

\bibitem{Dasgupta:2013cwa} 
  B.~Dasgupta, E.~Ma and K.~Tsumura,
  Phys.\ Rev.\ D {\bf 89}, 041702 (2014)
  [arXiv:1308.4138 [hep-ph]].

\bibitem{Hernandez:2013hea} 
  A.~E.~Carcamo Hernandez, RMartinez and F.~Ochoa,
  arXiv:1309.6567 [hep-ph].

\bibitem{McDonald:2013hsa} 
  K.~L.~McDonald,
  JHEP {\bf 1311}, 131 (2013)
  [arXiv:1310.0609 [hep-ph]].


\bibitem{Baek:2013fsa} 
  S.~Baek, H.~Okada and T.~Toma,
  JCAP {\bf 1406}, 027 (2014)
  [arXiv:1312.3761 [hep-ph]].
 

\bibitem{Ma:2014cfa} 
  E.~Ma,
  Phys.\ Lett.\ B {\bf 732}, 167 (2014)
  [arXiv:1401.3284 [hep-ph]].

\bibitem{Ahriche:2014xra} 
  A.~Ahriche, S.~Nasri and R.~Soualah,
  Phys.\ Rev.\ D {\bf 89}, 095010 (2014)
  [arXiv:1403.5694 [hep-ph]].

\bibitem{Okada:2014vla} 
  H.~Okada,
  arXiv:1404.0280 [hep-ph].
  

\bibitem{Ahriche:2014cda} 
  A.~Ahriche, C.~-S.~Chen, K.~L.~McDonald and S.~Nasri,
  arXiv:1404.2696 [hep-ph].
  
\bibitem{Ahriche:2014oda} 
  A.~Ahriche, K.~L.~McDonald and S.~Nasri,
  arXiv:1404.5917 [hep-ph].

\bibitem{Chen:2014ska} 
  C.~-S.~Chen, K.~L.~McDonald and S.~Nasri,
  Phys.\ Lett.\ B {\bf 734}, 388 (2014)
  [arXiv:1404.6033 [hep-ph]].

\bibitem{Kanemura:2014rpa} 
  S.~Kanemura, T.~Matsui and H.~Sugiyama,
  Phys.\ Rev.\ D {\bf 90}, 013001 (2014)
  [arXiv:1405.1935 [hep-ph]].

\bibitem{Aoki:2014cja} 
  M.~Aoki and T.~Toma,
  arXiv:1405.5870 [hep-ph].
  
\bibitem{Lindner:2014oea} 
  M.~Lindner, S.~Schmidt and J.~Smirnov,
  arXiv:1405.6204 [hep-ph].
  
  
 \bibitem{Ahn:2012cg} 
  Y.~H.~Ahn and H.~Okada,
  Phys.\ Rev.\ D {\bf 85}, 073010 (2012)
  [arXiv:1201.4436 [hep-ph]].

\bibitem{Ma:2012ez} 
  E.~Ma, A.~Natale and A.~Rashed,
  Int.\ J.\ Mod.\ Phys.\ A {\bf 27}, 1250134 (2012)
  [arXiv:1206.1570 [hep-ph]].
    
\bibitem{Kajiyama:2013lja} 
  Y.~Kajiyama, H.~Okada and K.~Yagyu,
  JHEP {\bf 10}, 196 (2013)
  arXiv:1307.0480 [hep-ph].
  
\bibitem{Kajiyama:2013sza} 
  Y.~Kajiyama, H.~Okada and K.~Yagyu,
  arXiv:1309.6234 [hep-ph].
  
\bibitem{Ma:2013mga} 
  E.~Ma,
  Phys.\ Rev.\ Lett.\  {\bf 112}, 091801 (2014)
  [arXiv:1311.3213 [hep-ph]].
  
\bibitem{Ma:2014eka} 
  E.~Ma and A.~Natale,
  Phys.\ Lett.\ B {\bf 723}, 403 (2014)
  [arXiv:1403.6772 [hep-ph]].


\bibitem{radlepton1} 
  H.~Okada and K.~Yagyu,
  Phys.\ Rev.\ D {\bf 89}, 053008 (2014)
  [arXiv:1311.4360 [hep-ph]].

\bibitem{radlepton2} 
  S.~Baek, H.~Okada and T.~Toma,
  Phys.\ Lett.\ B {\bf 732}, 85 (2014)
  [arXiv:1401.6921 [hep-ph]].

\bibitem{radlepton3} 
  H.~Okada and K.~Yagyu,
  arXiv:1405.2368 [hep-ph].



\bibitem{Schmidt:2014zoa} 
  D.~Schmidt, T.~Schwetz and H.~Zhang,
  arXiv:1402.2251 [hep-ph].

\bibitem{Long1} 
  H.~N.~Long and V.~V.~Vien,
  Int.\ J.\ Mod.\ Phys.\ A {\bf 29}, no. 13, 1450072 (2014)
  [arXiv:1405.1622 [hep-ph]].

\bibitem{Long2} 
  V.~Van Vien, H.~N.~Long and P.~N.~Thu,
  arXiv:1407.8286 [hep-ph].

\bibitem{Nebot} 
  J.~Herrero-Garcia, M.~Nebot, N.~Rius and A.~Santamaria,
  arXiv:1402.4491 [hep-ph].

\bibitem{discrepancy1}
  F.~Jegerlehner and A.~Nyffeler,
  Phys.\ Rept.\  {\bf 477}, 1 (2009)
  [arXiv:0902.3360 [hep-ph]].
  
\bibitem{discrepancy2} 
  M.~Benayoun, P.~David, L.~Delbuono and F.~Jegerlehner,
  Eur.\ Phys.\ J.\ C {\bf 72}, 1848 (2012)
 [arXiv:1106.1315 [hep-ph]].

\bibitem{Cheng:1980qt} 
  T.~P.~Cheng and L.~-F.~Li,
  Phys.\ Rev.\ D {\bf 22}, 2860 (1980).

\bibitem{Schechter:1980gr} 
  J.~Schechter and J.~W.~F.~Valle,
  Phys.\ Rev.\ D {\bf 22}, 2227 (1980).

\bibitem{Lazarides:1980nt} 
  G.~Lazarides, Q.~Shafi and C.~Wetterich,
  Nucl.\ Phys.\ B {\bf 181}, 287 (1981).

\bibitem{Mohapatra:1980yp} 
  R.~N.~Mohapatra and G.~Senjanovic,
  Phys.\ Rev.\ D {\bf 23}, 165 (1981).

\bibitem{Magg:1980ut} 
  M.~Magg and C.~Wetterich,
  Phys.\ Lett.\ B {\bf 94}, 61 (1980).



\bibitem{Ade:2013lta} 
  P.~A.~R.~Ade {\it et al.}  [ Planck Collaboration],
  arXiv:1303.5076 [astro-ph.CO].

\bibitem{Hambye:2009pw} 
  T.~Hambye, F.~-S.~Ling, L.~Lopez Honorez and J.~Rocher,
  JHEP {\bf 0907}, 090 (2009)
  [Erratum-ibid.\  {\bf 1005}, 066 (2010)]
  [arXiv:0903.4010 [hep-ph]].

\bibitem{LopezHonorez:2006gr} 
  L.~Lopez Honorez, E.~Nezri, J.~F.~Oliver and M.~H.~G.~Tytgat,
  JCAP {\bf 0702}, 028 (2007)
  [hep-ph/0612275].

\bibitem{Akerib:2013tjd} 
  D.~S.~Akerib {\it et al.}  [LUX Collaboration],
  arXiv:1310.8214 [astro-ph.CO].



\bibitem{TuckerSmith:2001hy} 
  D.~Tucker-Smith and N.~Weiner,
  Phys.\ Rev.\ D {\bf 64}, 043502 (2001)
  [hep-ph/0101138].



\bibitem{stu}
  M.~E.~Peskin and T.~Takeuchi,
  Phys.\ Rev.\ Lett.\  {\bf 65}, 964 (1990); 
  M.~E.~Peskin and T.~Takeuchi, 
  Phys.\ Rev.\  D {\bf 46}, 381 (1992). 
  
  \bibitem{Baak:2012kk}
  M.~Baak, M.~Goebel, J.~Haller, A.~Hoecker, D.~Kennedy, R.~Kogler, K.~Moenig and M.~Schott {\it et al.},
  Eur.\ Phys.\ J.\ C {\bf 72}, 2205 (2012)
  [arXiv:1209.2716 [hep-ph]].


\bibitem{AristizabalSierra:2006gb} 
  D.~Aristizabal Sierra and M.~Hirsch,
  JHEP {\bf 0612}, 052 (2006)
  [hep-ph/0609307].

\bibitem{pmns} 
  Z.~Maki, M.~Nakagawa and S.~Sakata,
  Prog.\ Theor.\ Phys.\  {\bf 28}, 870 (1962).


\bibitem{neutrino_mixing} 
  D.~V.~Forero, M.~Tortola and J.~W.~F.~Valle,
  arXiv:1405.7540 [hep-ph].

\bibitem{neutrino_sum} 
  F.~Beutler {\it et al.}  [BOSS Collaboration],
  arXiv:1403.4599 [astro-ph.CO].


\bibitem{bennett}
  G.~W.~Bennett {\it et al.}  [Muon G-2 Collaboration],
  Phys.\ Rev.\ D {\bf 73}, 072003 (2006)
  [hep-ex/0602035].

 

\bibitem{slepton} 

 [ATLAS Collaboration], 
  Report No. ATLAS-CONF-2013-049. 

\bibitem{calchep}
  A.~Pukhov,
  [hep-ph/0412191].




\bibitem{Babu:2002uu} 
  K.~S.~Babu and C.~Macesanu,
  Phys.\ Rev.\ D {\bf 67}, 073010 (2003)
  [hep-ph/0212058].


\bibitem{Nebot:2007bc} 
  M.~Nebot, J.~F.~Oliver, D.~Palao and A.~Santamaria,
  Phys.\ Rev.\ D {\bf 77}, 093013 (2008)
  [arXiv:0711.0483 [hep-ph]].






\bibitem{Perez:2008ha} 
  P.~Fileviez Perez, T.~Han, G.~-y.~Huang, T.~Li and K.~Wang,
  Phys.\ Rev.\ D {\bf 78}, 015018 (2008)
  [arXiv:0805.3536 [hep-ph]].

\bibitem{Chiang:2012dk} 
  C.~-W.~Chiang, T.~Nomura and K.~Tsumura,
  Phys.\ Rev.\ D {\bf 85}, 095023 (2012)
  [arXiv:1202.2014 [hep-ph]].

\bibitem{Kanemura:2013vxa} 
  S.~Kanemura, K.~Yagyu and H.~Yokoya,
  Phys.\ Lett.\ B {\bf 726}, 316 (2013)
  [arXiv:1305.2383 [hep-ph]].

\bibitem{kang:2014jia} 
  Z.~Kang, J.~Li, T.~Li, Y.~Liu and G.~-Z.~Ning,
  arXiv:1404.5207 [hep-ph].

\bibitem{Kanemura:2014goa} 
  S.~Kanemura, M.~Kikuchi, K.~Yagyu and H.~Yokoya,
  arXiv:1407.6547 [hep-ph].




\end{thebibliography}
\end{document}